\documentclass[sigconf, nonacm, balance=false, prologue, table]{acmart}
\acmSubmissionID{535}

\usepackage{dsfont}
\usepackage{wrapfig}
\usepackage{xspace}
\usepackage{amsmath}
\usepackage{mymacros}
\usepackage{xcolor}
\usepackage{multirow}
\usepackage{empheq}
\usepackage{mathtools}
\usepackage{float}
\AtBeginDocument{%
  \providecommand\BibTeX{{%
    \normalfont B\kern-0.5em{\scshape i\kern-0.25em b}\kern-0.8em\TeX}}}
    
\definecolor{lightgray}{gray}{0.9}
\copyrightyear{2023}
\acmYear{2023}
\setcopyright{acmlicensed}\acmConference[SA Conference Papers '23]{SIGGRAPH Asia 2023 Conference Papers}{December 12--15, 2023}{Sydney, NSW, Australia}
\acmBooktitle{SIGGRAPH Asia 2023 Conference Papers (SA Conference Papers '23), December 12--15, 2023, Sydney, NSW, Australia}
\acmPrice{15.00}
\acmDOI{10.1145/3610548.3618192}
\acmISBN{979-8-4007-0315-7/23/12}
\begin{document}

%%
%% The "title" command has an optional parameter,
%% allowing the author to define a "short title" to be used in page headers.
\title{Explorable Mesh Deformation Subspaces from Unstructured 3D Generative Models}

%%
%% The "author" command and its associated commands are used to define
%% the authors and their affiliations.
%% Of note is the shared affiliation of the first two authors, and the
%% "authornote" and "authornotemark" commands
%% used to denote shared contribution to the research.
\author{Arman Maesumi}
\email{arman_maesumi@brown.edu}
\affiliation{%
  \institution{Brown University}
  \city{Providence}
  \country{USA}
}

\author{Paul Guerrero}
\email{guerrero@adobe.com}
\affiliation{%
  \institution{Adobe Research}
  \city{London}
  \country{UK}
}

\author{Vladimir G. Kim}
\email{vokim@adobe.com}
\affiliation{%
 \institution{Adobe Research}
 \city{Seattle}
 \country{USA}
}

\author{Matthew Fisher}
\email{mdfisher@cs.stanford.edu}
\affiliation{%
  \institution{Adobe Research}
  \city{San Francisco}
  \country{USA}
}

\author{Siddhartha Chaudhuri}
\email{sidch@adobe.com}
\affiliation{%
  \institution{Adobe Research}
  \city{New York}
  \country{USA}
}

\author{Noam Aigerman}
\email{noam.aigerman@umontreal.ca}
\affiliation{%
  \institution{University of Montreal\\Adobe Research}
  \city{Montreal}
  \country{Canada}
}

\author{Daniel Ritchie}
\email{daniel_ritchie@brown.edu}
\affiliation{%
  \institution{Brown University}
  \city{Providence}
  \country{USA}}

%%
%% By default, the full list of authors will be used in the page
%% headers. Often, this list is too long, and will overlap
%% other information printed in the page headers. This command allows
%% the author to define a more concise list
%% of authors' names for this purpose.
\renewcommand{\shortauthors}{Maesumi et al.}

%%
%% The abstract is a short summary of the work to be presented in the
%% article.
\begin{abstract}
Exploring variations of 3D shapes is a time-consuming process in traditional 3D modeling tools. Deep generative models of 3D shapes often feature continuous latent spaces that can, in principle, be used to explore potential variations starting from a set of input shapes; in practice, doing so can be problematic---latent spaces are high dimensional and hard to visualize, contain shapes that are not relevant to the input shapes, and linear paths through them often lead to sub-optimal shape transitions.
Furthermore, one would ideally be able to explore variations in the original high-quality meshes used to train the generative model, not its lower-quality output geometry.
In this paper, we present a method to explore variations among a given set of \emph{landmark} shapes by constructing a mapping from an easily-navigable 2D exploration space to a subspace of a pre-trained generative model.
We first describe how to find a mapping that
spans the set of input landmark shapes and exhibits smooth variations between them.
We then show how to turn the variations in this subspace into deformation fields, to transfer those variations to high-quality meshes for the landmark shapes.
Our results show that our method can produce visually-pleasing and easily-navigable 2D exploration spaces for several different shape categories, especially as compared to prior work on learning deformation spaces for 3D shapes.

\noindent\href{https://github.com/ArmanMaesumi/generative-mesh-subspaces}{\textcolor{blue}{https://github.com/ArmanMaesumi/generative-mesh-subspaces}}
\end{abstract}

%%
%% The code below is generated by the tool at http://dl.acm.org/ccs.cfm.
%% Please copy and paste the code instead of the example below.
%%
\begin{CCSXML}
\end{CCSXML}
\begin{CCSXML}
<ccs2012>
   <concept>
       <concept_id>10010147.10010371.10010396.10010402</concept_id>
       <concept_desc>Computing methodologies~Shape analysis</concept_desc>
       <concept_significance>500</concept_significance>
       </concept>
 </ccs2012>
\end{CCSXML}
\ccsdesc[500]{Computing methodologies~Shape analysis}

%%
%% Keywords. The author(s) should pick words that accurately describe
%% the work being presented. Separate the keywords with commas.
\keywords{shape deformation, generative model, 3D shape generation}

%% A "teaser" image appears between the author and affiliation
%% information and the body of the document, and typically spans the
%% page.
\begin{teaserfigure}
  \includegraphics[width=\textwidth, trim=0 10px 0 0]{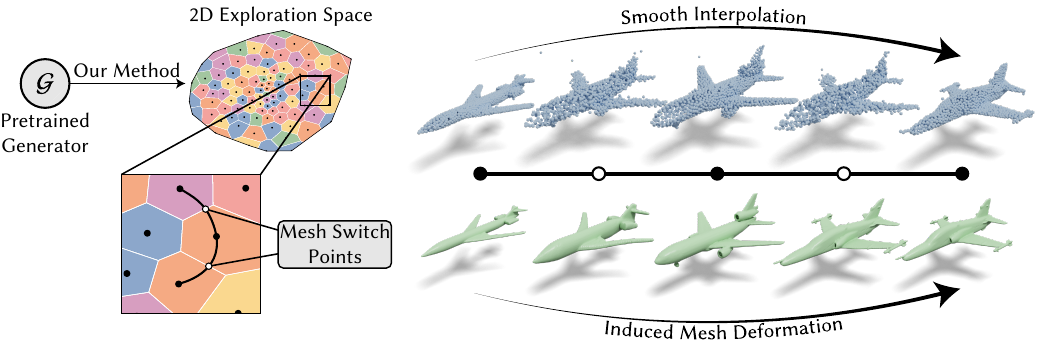}
  \caption{We present a method to explore variations among a given set of input shapes (denoted by black Voronoi centers on the left)
  using a two-dimensional exploration space. This exploration space smoothly and naturally interpolates between the input shapes by constructing a mapping to a sub-space of a pre-trained generator's latent space that optimizes the smoothness of interpolations along any trajectory. Additionally, we transfer the variation over these interpolation trajectories onto the original high-quality meshes, avoiding loss of detail from the unstructured generator output.}
  \label{fig:teaser}
\end{teaserfigure}

% \received{20 February 2007}
% \received[revised]{12 March 2009}
% \received[accepted]{5 June 2009}

%%
%% This command processes the author and affiliation and title
%% information and builds the first part of the formatted document.
\maketitle
\section{Introduction}
When designing a 3D shape, artists, designers, or engineers may want to explore variations of initial designs, to find shapes with better aesthetics or functional properties. In traditional 3D modelling tools, this exploration is time-consuming, as each variation needs to created manually.
Data-driven generative models have introduced revolutionary new capabilities to the practice of 3D shape design and exploration.
This trend began over a decade ago with methods for generating 3D shapes by recombining parts from existing shapes~\cite{chaudhuri2011probabilistic,kalogerakis2012,chaudhuri2013attribit}.
With the emergence of deep learning, the movement has shifted to deep generative models~\cite{3dgan,pointflow,li2021spgan,chen2018imnet,Achlioptas2017LearningRA,zheng2022sdfstylegan}.
These latter methods are particularly transformative: in addition to synthesizing new shapes, these models often feature \emph{latent spaces} that allow navigation through a continuous manifold of shapes. 

In principle, these latent spaces are useful to explore variations of initial designs.  
However, latent spaces have some limitations when used for this purpose:
First, latent spaces contain the full distribution of shapes that the generative model was trained on, but we want to focus on variations among a small set of initial shapes of interest.
Second, while individual points in the latent space typically correspond to plausible shapes, following a linear path between two such points typically does not produce the most natural transition between the shapes they represent---often, extraneous shape variations and/or noise occur along the way.
Third, these latent spaces are typically high dimensional (e.g. $\Reals^{128}$ or higher), making them hard to visualize and interact with.
Such spaces may be acceptable for goal-directed interpolation between two shapes, but they remain difficult to use for open-ended, interactive exploration.
Finally, the best state-of-the-art generative models output point clouds or implicit fields~\cite{li2021spgan,zheng2022sdfstylegan} to handle the challenge of modeling shape distributions with varying topologies, yet many downstream graphics applications demand mesh representations.
While point clouds and implicit fields can be meshed, the resulting meshes are not as high-quality as those designed by expert artists.

In this paper, we propose a new method for exploring continuous variations among a set of given \emph{landmark} meshes, along with discrete transitions between them.
Rather than trying to explore the full latent space of a pre-trained generative model, or finding an embedding that preserves latent space distances, we extract a smaller shape space that smoothly spans the landmark. This reduced space is used as a prior for downstream tasks (i.e. shape deformation).

Our method consists of two main components. First, we develop a technique to construct a mapping from an easily-navigable \emph{exploration space} to a subspace of a pre-trained shape generative model.
Given a set of landmark shapes (which may be from the generative model's training set, or another set of shapes of the same category), we embed these shapes in two dimensions to facilitate interaction.
We then learn a map from this 2D exploration space to the generative model's latent space, such that (a) the embedded points map to latent space points that produce their corresponding landmark shapes when put through the pre-trained generator, and importantly (b) navigating between landmark points produces smooth shape variations, rather than preserving latent space distances.

Second, we show how to use this exploration space to explore variations within and between the high-quality original \emph{meshes} for the landmark shapes.
Specifically, we develop a technique for computing continuous deformation fields from small steps within the exploration space, allowing its rich semantic variations to be transferred to meshes at real-time interactive rates.
We also consider how and when to switch between deformed landmark meshes as the user navigates the 2D space to minimize jarring visual discontinuities.

We evaluate our method by producing easily-navigable and visually-pleasing exploration spaces for several different shape categories.
We compare the paths through latent space that our method produces to those produced by alternative approaches, including a method for directly learning a deformation space from a large collection of shapes~\cite{jiang2020shapeflow}.
We also present a set of ablations and analyses of our method's components.

In summary, our contributions are:
\begin{packed_itemize}
    \item A technique for building a 2D exploration space that naturally interpolates between a set of landmark shapes while staying on the shape manifold induced by a pre-trained generative model.
    \item A method for transferring latent space variations into continuous deformations of high-quality landmark meshes and discrete transitions between them in real-time (accompanied by a graphical user interface).
\end{packed_itemize}

\section{Related work}
\begin{figure}
    \centering
    \includegraphics[width=\linewidth]{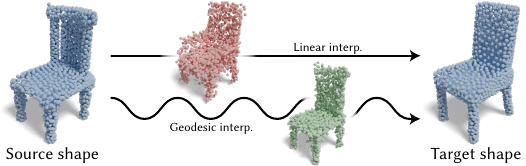}
    \caption{Linearly interpolating between shapes in latent space may produce poor intermediate samples. Here we see the difference between linear and non-linear (geodesic) interpolation at $t=0.5$ in SP-GAN's latent space. The linear interpolation is noisy and is beginning to grow armrests, despite the source and target shapes lacking such features.}
    \label{fig:linearVsPoly}
    \vspace{-1em}
\end{figure}
% \emph{Deep generative models of 3D shapes}
\paragraph{Deep generative models of 3D shapes}
There has been an explosion of work in recent years on applying deep generative models to the problem of synthesizing 3D shapes.
In this paper, we are concerned with \emph{latent variable deep generative models}, i.e. generative models that map points in a latent space $\mathcal{Z}$ to the output shape domain via some map $f$. A variety of such models exist, including variational autoencoders (VAEs)~\cite{li2017grass,jones2020SA}, generative adversarial networks (GANs) ~\cite{3dgan,chen2018implicit_decoder,Achlioptas2017LearningRA,zheng2022sdfstylegan}, normalizing flows~\cite{pointflow}, and denoising diffusion probabilistic models (DDPMs)~\cite{zeng2022lion,NeuralWaveletDiffusion}.
Our method is designed to work well with such generative models in which $f$ is a single forward pass through a neural network (e.g. VAEs, GANs).
We discuss challenges with, and potential ideas for, applying it to multi-step $f$ models (e.g. flows, DDPMs) in Section~\ref{sec:conclusion}.
Our method is not designed to work with autoregressive models, which generate outputs through a sequence of sampling steps~\cite{autosdf2022}.

There have also been attempts at learning generative models that output meshes directly.
Thus far, these have suffered from various limitations: restriction to genus zero topology~\cite{wang2018pixel2mesh}, requiring part-segmented data~\cite{gao2019sdmnet,Jie20DsgNet}, or unpredictable output quality~\cite{nash2020polygen}.
Rather than \emph{generate} meshes directly, our approach allows exploration of variations between existing high-quality meshes by borrowing the shape variations exhibited by a pre-trained non-mesh generator.

\paragraph{Exploring collections of 3D shapes}
Several prior techniques have been proposed for exploring collections of 3D shapes by deforming a template~\cite{maks_exploration},  highlighting regions on some reference shapes~\cite{Kim12}, or varying manually annotated language-based attributes~\cite{chaudhuri2013attribit}. These interfaces typically focus on modifying a single reference shape, and thus are more suitable for local search, exploring immediate neighbors of the reference model. A global analysis has been proposed to identify main modes of variations of shapes in a collection~\cite{Kim13}, and 2D layouts are commonly used to jointly explore shape collections or generative shape spaces~\cite{Averkiou14, ExplorationInverseMapping}, as well as other visual domains in general ~\cite{DynamicMaps, ExploratoryModeling, marks1997design}. These type of 2D layouts give a holistic view of the entire space, while still providing intuitive interfaces in 2D. Various techniques have been developed in mapping high-dimensional similarity data to 2D (e.g., PCA, MDS, ~\cite{Fried:2015:ICI}, ~\cite{Dym17}) for exploration and visualization. In this work we propose a novel technique for constructing 2D exploration space from arbitrary high-dimensional latent spaces learned with some shape generation techniques.

\paragraph{Learning to deform 3D shapes}
Classical methods typically define shape deformation in terms of optimizing a physical or quasi-physical energy, e.g. see the survey by Botsch and Sorkine~\shortcite{botsch2008survey}. These methods, while mathematically well-founded and tied to actual physical behavior, can be difficult to control, do not capture non-physical effects (e.g. interpolating in a heterogeneous shape space as studied in this paper) and frequently involve numerically challenging optimization (sometimes bypassable with ``forward'' models like skinning, e.g.~\cite{jacobson2011bbw}). To address these issues, a number of papers have tried to apply neural networks to learn deformation models in a data-driven way. Here, we discuss a few representative ones. To address the challenge of smoothly deforming complex geometry, several papers use machine learning to infer classical low-dimensional controls. Yifan et al.~\shortcite{yifan2020neuralcages} train a network to fit and deform low-dimensional cages for 3D shapes. Liu et al.~\shortcite{liu2021metahandles} take a different approach, learning combinations of shape control points that factorize the deformation space into intuitive ``meta-handles''. Xu et al.~\shortcite{xu2022morig} learn to infer animation rigs for meshes from motion-captured point cloud sequences. In contrast, Aigerman et al.~\shortcite{aigerman2022njf} learn an implicit Jacobian field that deforms meshes to targets {\em without} an intermediate classical proxy. While all of these methods provide interesting ways to deform or interpolate between meshes, they are not well suited for shape space exploration, which is the problem we study in this paper.

A work more directly related to ours is the ShapeFlow system of Jiang et al.~\shortcite{jiang2020shapeflow}. This system learns, in parallel, neural flow fields that deform one shape to another, as well as an embedding of all training shapes to a latent space. In principle, ShapeFlow can be applied to our problem, and we show in the evaluation section that our method produces qualitatively better deformations, see Figure \ref{fig:shapeFlowComparison}. However, beyond that, ShapeFlow does not optimize the latent space for interactive two-dimensional exploration -- thus, much of the machinery we develop for extracting a human-friendly 2D re-embedding of the original latent space would be necessary in any case. Secondly, our method can leverage any arbitrary pre-trained generative 3D model as a backbone, including ones trained on much larger 3D (or 2D) datasets than the collection being explored. 

A method that does use latent-space generative models for shape space exploration is the GLASS system of Muralikrishnan et al.~\shortcite{muralikrishnan2022glass}. However, the problem addressed by this paper is very different from ours -- they focus on discovering new deformations of a single template mesh by alternately training a generative model, and incrementally exploring its latent space guided by a pre-defined physical energy. This method does not apply to heterogeneous shape collections, nor does it address interactive exploration.

% \vspace{-3em}
\section{Overview}
\begin{figure*}[hbt!]
    \centering
    \includegraphics[width=\linewidth,trim=0 1em 0 0]{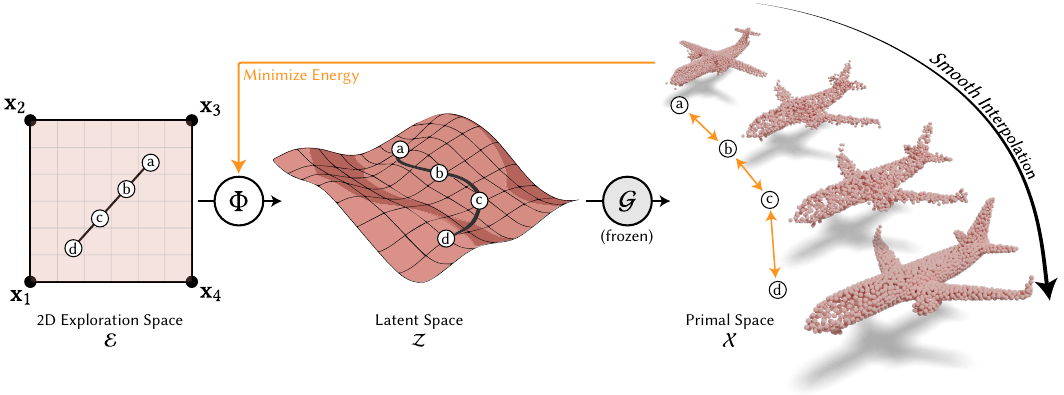}
    \caption{We illustrate the effect of our energy-minimizing submanifold in a simple case with four landmarks in exploration space. The region inscribed by the landmarks is mapped to $\G$'s latent space via $\map$, as shown by the four interior points $a,b,c,d$. These points are decoded by the generator, producing point clouds of a particular shape category. The energy over these point clouds is said to be minimized by $\map$ through the optimization outlined in Section \ref{sec:submanifold}.}
    \label{fig:submanifoldPipeline}
\end{figure*}
\label{sec:overview}
Given a pre-trained, unstructured shape generative model $\G$ (e.g. a point cloud generator) and a set of \emph{landmark meshes} $\meshes = \{M_1, ..., M_N\}$ from the same shape category of which $\G$ was trained, our goal is to create a two-dimensional \emph{exploration space}, $\explore$, that can be used to navigate a deformation subspace over the landmark meshes induced by $\G$. Walking through exploration space should produce smoothly varying and sensible deformations of the meshes, and the space's layout should be such that similar meshes in $\meshes$ are embedded closer together, while dissimilar meshes are farther apart.

We define the exploration space via a function $\map$ that maps from $\explore$ to the generator's latent space $\Z$. The mapping is designed to give the exploration space several desirable properties that the generator lacks; for instance, interpolating between shapes in $\explore$ produces intermediate outputs that vary smoothly, whereas naive linear interpolation through most generative models may produce samples that contain excessive variation, as illustrated in Figure \ref{fig:linearVsPoly}. At a high level, $\map$ reparametrizes $\Z$ in a way that minimizes energy in the \emph{primal space} of the generator (its geometric output space), thus avoiding such issues. Additionally, $\map$ allows the region of $\Z$ spanned by our given meshes to be visualized in two dimensions, making the high-dimensional latent space easily navigable. 

Armed with $\map$, there is one undesirable property remaining: $\explore$ only permits exploration over the unstructured outputs of $\G$ (e.g. point clouds), rather than variations of the high-quality input meshes.
To solve this problem, we introduce a deformation module that interprets an interpolation through the latent space of $\G$ as a flow on a mesh's vertices, producing detail-preserving deformations of a wide variety of shapes.
We show how to advect the input meshes through this flow, as well as how to switch between which mesh is being advected based on a partitioning of the 2D space.

In the following two sections of the paper, we first outline the construction of the energy-minimizing map $\map$ into the generator's latent space. Then, we introduce our routine that transforms the unstructured shape manifold into a mesh deformation space.
% \begin{figure*}[hbt!]
%     \centering
%     \includegraphics[width=\linewidth,trim=0 1em 0 0]{figs/fig_pipeline.pdf}
%     \caption{We illustrate the effect of our energy-minimizing submanifold in a simple case with four landmarks in exploration space. The region inscribed by the landmarks is mapped to $\G$'s latent space via $\map$, as shown by the four interior points $a,b,c,d$. These points are decoded by the generator, producing point clouds of a particular shape category. The energy over these point clouds is said to be minimized by $\map$ through the optimization outlined in Section \ref{sec:submanifold}.}
%     \label{fig:submanifoldPipeline}
%     \vspace{-1em}
% \end{figure*}
\section{Constructing the 2D Exploration Space 
$\explore$}
\label{sec:2dspace}
We now detail the inner workings of our exploration space. We assume to be given a \emph{shape manifold}, i.e., a continuous space in which each point represents a shape --- this space can be thought of as a submanifold of primal space (the space of all possible shapes). We consider a shape manifold defined by a pretrained generative neural network, $\G$. Concretely, such generators indeed define a $k$-manifold, if one considers it as a function taking us from latent (Euclidean) space, $\Z \in \mathbb{R}^k$, onto points in primal space $\G(\*z) \subset \X$.

Our goal is to extract a submanifold from the shape manifold that 1) includes all of the given landmark shapes; 2) naturally interpolates between the landmarks along the shape manifold 3) is a $2$-manifold, i.e., is parameterized via a map  $\map$, which maps the two-dimensional space into the shape manifold, $\map : \explore \subset \mathbb{R}^2 \mapsto \Z$. 

Say we have a two-dimensional embedding of our landmark meshes (detailed in Section \ref{sec:embeddingMeshes}) that will act as our parametrization of the aforementioned map, i.e. for each mesh we associate with it a point in exploration space (2D) and its corresponding projection into latent space: $\*L = \{(\*x_1, \*z_1), ..., (\*x_N, \*z_N)\}$. As illustrated in Figure \ref{fig:submanifoldPipeline} we seek a map that ``naturally'' lifts this exploration space into the shape manifold, i.e., in a manner that does not exhibit excessive variation or rapid changes in the features of the shapes.

We can formalize this requirement: the shape manifold may be defined in the Euclidean space $\Z$, however the shapes themselves are generated by a non-linear mapping (the generator), thus small, linear steps in $\Z$ may result in arbitrarily large changes in the resulting shape. In other words, we need to account for the \emph{metric} of the shape manifold and the curvature it induces. We will require that the map $\map$ is smooth w.r.t the metric of the shape manifold.

By casting this as an embedding problem, we formulate a constrained optimization objective that ensures the landmark shapes are always mapped correctly and the embedding minimizes an energy accounting for the metric of the shape manifold -- the Dirichlet energy \cite{karcher}. We first discuss this energy in the context of one-dimensional interpolation problems. Then, the two-dimensional problem is laid out for a simple case where just three landmarks are specified in Section \ref{sec:submanifold}. Finally, we generalize our method for collections of scattered landmarks in Section \ref{sec:scatteredLandmarks}.

\subsection{Geodesic paths between landmarks}
Linear interpolation in high-dimensional latent spaces is known to produce samples that exhibit undesirable properties \citep{stylegan2018, laine2018feature-based}. Intermediate samples may contain features that do not exist at either the source or target; similarly, $\G$ may contain regions with excessive variation, resulting in a ``jittery'' output.

In order to avoid such interpolants, we optimize for paths in latent space that are \emph{geodesics} with respect to primal space, $\X$. This is done by finding paths, $p(t)$, $t \in [0,1]$, where $p(0)=\*z_0$, $p(1)=\*z_1$, that minimize the \emph{Dirichlet energy} in primal space
\begin{equation}
\textstyle
    \J_{\,\text{1-D}}(p) = \frac{1}{2}\int_0^1 \|\nabla \G (p(t))\|^2 dt.
    \label{eq:1DDirichlet}
\end{equation}
As geodesics are (locally) shortest paths, the shapes along them vary as gradually as possible, and thus do not contain superfluous movement in the generated samples. Figure \ref{fig:linearVsPoly} illustrates the difference between linear and geodesic interpolation through $\Z$.

These geodesics are useful for smoothly interpolating between samples along a one dimensional path, but they provide no information about smoothly interpolating through a solid region in latent space. In the simplest case, imagine three shapes connected by geodesic paths; we have a well-defined way to walk directly from shape to shape, but this formulation does not tell us how to walk to an arbitrary point in the \textit{interior} of the region.
% -------------------------------------------------- %
\subsection{Energy-minimizing submanifolds of $\G$}\label{sec:submanifold}

We now formulate the constrained optimization problem in a case where three landmarks $\*z_1$, $\*z_2$, and $\*z_3$ are specified. Then, our solution is generalized in Section \ref{sec:scatteredLandmarks}. Since geodesic paths between any pair of landmarks map to straight lines in exploration space $\explore$, the three landmarks and the geodesic paths connecting them, $p_{12}(t)$, $p_{23}(t)$, $p_{31}(t)$ form the vertices $\*x_1$, $\*x_2$, $\*x_3$ and edges $\Vec{\*x}_{12}$, $\Vec{\*x}_{23}$, $\Vec{\*x}_{31}$ of a triangle in $\explore$. We then define the solution surface $\surface(x) = \G(\Phi(\*x))$ with $\*x \in \explore$ inside the triangle as the minimizer of the Dirichlet energy w.r.t to the Jacobian $\mathbf{J}$ of $\G\circ\map$ over exploration space:
% \vspace{-0.1em}
  \begin{empheq}[box=\Garybox]{gather}
    \nonumber\argmin_\Phi\quad \dfrac{1}{2}\int_\explore \|\, \mathbf{J}(\*x)\|^2_F \, dA,\\
    \notag\text{\parbox{\linewidth}{\, subject to a Dirichlet boundary condition on $\partial \explore$ given by $\Vec{\*x}_{ij}$}}\\
    \nonumber\Phi(\*x) = b(\*x) \text{ if } \*x \in \Vec{\*x}_{ij} \\
    \nonumber\text{ with } b(\*x) = p_{ij}\Big(\frac{\|\*x - \*x_i\|}{\|\*x_j - \*x_i\|}\Big)
  \end{empheq}
In order to optimize for $\map$, we start by discretizing the function $\G(\map(\*x))$ on a dense finite-element triangulation $(\mathbf{V}, \mathbf{F})$ over the exploration space, with vertices $\mathbf{V}$ and faces $\mathbf{F}$. Details of this discretization can be found in Section \ref{sec:implementation}. Assuming the discretized function is piecewise linear on each triangle, the Jacobian $\mathbf{J}$ is piecewise constant, and we denote the Jacobian for each triangle $f_i \in \mathbf{F}$ as $\mathbf{J}^i$. The discretized objective function for our optimization is then
\begin{equation}
    \mathcal{J}_{\,\text{2-D}} = \sum_{f_i \in \*F} \|\, \*J^i \|^2_F \cdot \area(f_i).
    \label{eq:finalEnergy}
\end{equation}

In order to enforce our boundary condition, we could add a point-matching regularizer in similar spirit to the thin plate spline energy; thus making the energy
\begin{displaymath}
    \mathcal{J}_{\,\text{reg}} = \sum_{f_i \in \*F} \|\, \*J^i \|_F^2 \cdot \area(f_i) + \underbrace{\lambda \int_{\partial\explore} \|b(\*x)  - \Phi(\*x)\|^2 dA}_{\text{unstable soft constraint}}.
\end{displaymath}
This energy, however, is unstable in practice due to the regularizer acting as a \emph{soft} constraint. The choice of $\lambda$ may catastrophically cause the solution surface to collapse to a point, or cause the surface to have regions with large gradients. Instead, we treat $\Phi(\*x)$ as a semi-parametric function, where its functional value is replaced with the corresponding boundary value $b(\*x)$ if $\*x \in \partial \explore$, thereby ``pinning'' the boundary points in place, turning them into a \textit{hard} constraint. The remaining regions of the surface are represented by a multi-layer perceptron parametrized by $\theta$:

\begin{equation}
    \Phi(\*x) = 
    \begin{cases}
    b(\*x) &\text{if } \*x \in \partial \explore\\%\partial \explore \\
    \text{MLP}_\theta(\*x) &\text{otherwise.}
    \end{cases}
    \label{eq:piecewiseMLP}
\end{equation}
In summary, we resolve the optimal mapping by minimizing the energy given by Equation $\ref{eq:finalEnergy}$ w.r.t. the semi-parametric MLP above.

% -------------------------------------------------------- %
\subsection{Generalizing to scattered landmarks in $\explore$} \label{sec:scatteredLandmarks}
\newcommand{\tentpole}{
\setlength{\columnsep}{1em}
\setlength{\intextsep}{0em}
\begin{wrapfigure}[7]{r}{70pt}
    \centering
    \includegraphics[width=\linewidth, trim=0 0 0 25px]{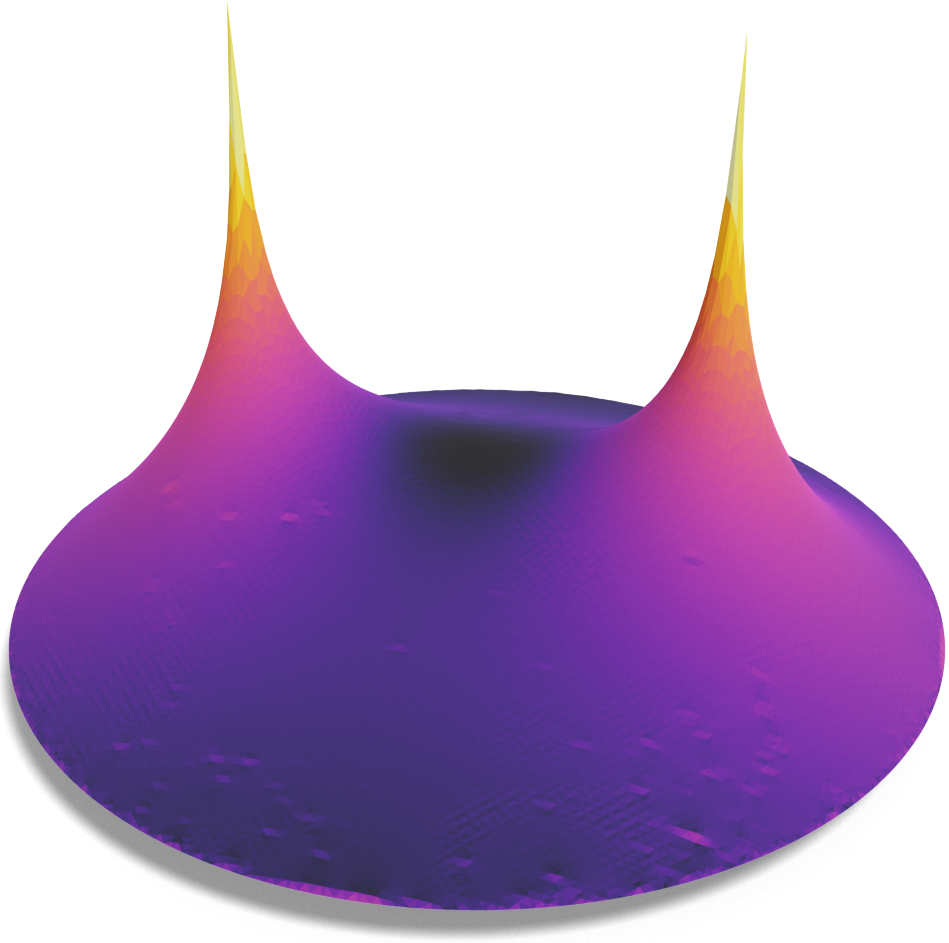}    
\end{wrapfigure}
}
Rather than operating on just three landmarks, our method facilitates exploration across an entire collection of shapes by defining $\map$ as a map from a convex region in $\explore$ populated by a set of scattered landmarks $\*L$. However, Laplace's equation (which is implicitly being solved by our energy) and similar PDEs are ill-defined with scattered boundary conditions\footnote{See anonymous StackExchange answer \shortcite{tentpoleSE} for further details.}.

\tentpole{}For instance, if we take our boundary condition to be, say, a set of values along the convex hull of $\explore$ along with the points in $\*L$, then the solution surface $\surface$ will exhibit a ``tentpole'' effect---areas with large gradients will protrude from the surface, like poles poking through a circus tent (see inset figure).

Instead, we partition $\explore$ into triangular facets via a Delaunay triangulation of the landmarks. This creates disjoint regions, which can be thought of as separate interpolation problems; that is, we can imagine $\surface$ as being defined by a collection of neighboring surface patches that are ``stiched'' together. So long as each patch shares identical boundary conditions with its neighbors, then $\surface$ will be continuous. We map each edge $(\*x_i, \*x_j)$ in the triangulation to a corresponding geodesic connecting $\*z_i$ to $\*z_j$. Hence our full boundary condition becomes $\text{lerp}(\*x_i, \*x_j) \hookrightarrow \text{geodesic}(\*z_i, \*z_j)$ for all edges in the triangulation, where lerp is linear interpolation. 

\subsection{Embedding meshes into $\explore$}\label{sec:embeddingMeshes}
To initially embed our meshes into $\explore$, we create a k-nearest neighbor graph of the meshes based on a suitable distance metric (specified in Section \ref{sec:implementation}). This graph is embedded into $\explore$ using a triplet margin loss \cite{tripletLoss}, where triplets of meshes (a tuple comprised of an anchor, positive neighboring mesh, and negative non-neighbor) are used to minimize distances between the anchors and their neighbors, while maximizing their distance to the remaining shapes. The triplet loss over all such tuples is given as
\begin{displaymath}
    \mathcal{L}_{\mathrm{triplet}}(a, p, n) = \sum_i^{k\cdot N} \max(\|a_i - p_i\|^2_2 - \|a_i - n_i\|^2_2 + \alpha, 0)
\end{displaymath}
where $\alpha$ is the margin parameter that controls the separation between positive and negative pairs. We note that the choice of triplet loss is arbitrary (methods such as t-SNE or UMAP may be used as well \cite{TSNE,UMAP}); however, in our experiments we found the embeddings produced by the triplet loss contained sufficient local similarity between landmarks.

We include additional regularizer terms that maintain uniformity in spacing between the embedded points. In particular, we penalize areas and interior angles of triangles induced by a Delaunay triangulation of the embedded points via $\ell^2$ norms of the areas and minimum interior angles of each triangle. Avoiding a poor triangulation of these points is pertinent to the boundary conditions established in Section \ref{sec:submanifold}.

\section{Explorable mesh deformation subspaces}\label{sec:deformation}
We now have an explorable subspace---parametrized by our two-dimensional energy-minimizing map $\map$---in $\Z$ that spans a set of landmark shapes. Instead of simply synthesizing unstructured shapes from $\G$ (i.e. point clouds), we would like to produce high quality \emph{meshes} that vary smoothly as we walk through the subspace. In the following sections, we introduce a novel method for transforming $\explore$ into a mesh deformation subspace. Then, we generalize this method to accommodate meshes with varying topology.

\subsection{Transforming $\explore$ into a mesh deformation subspace}
Given a landmark point in exploration space, $\*x_{\text{src}} \in \explore$, and its corresponding mesh $\*M (\*V, \*F) \in \meshes$, how can we deform the mesh to match an arbitrary point $\*x_{\text{tar}}$ whose structure is given by $\surface(\*x_{\text{tar}})$? The key to our deformation module is to interpret an interpolation through exploration space as a flow on the mesh's vertices, $\*V$.

\begin{figure}
    \centering
    \includegraphics[width=\linewidth, trim=0.25cm 0.25cm 0.25cm 0]{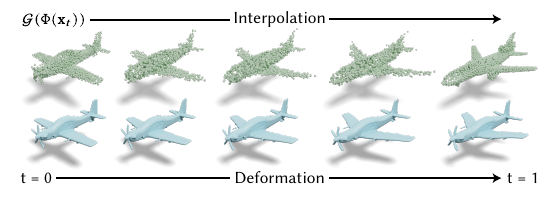}
    \caption{We exploit $\G$'s unstructured shape manifold to produce plausible deformations of meshes. Our deformation module is driven by interpolations through our map $\G(\map(\*x))$, which offers smoother variation through exploration space.}
    \label{fig:exampleDeform}
    % \vspace{-2em}
\end{figure}

More concretely, say we have a continuous path on our submanifold, $\*S = (\surface(\*x_0), ..., \surface(\*x_t) \mid \forall t \in [0, 1])$ where $\*x_0 = \*x_{\text{src}}$, $\*x_1 = \*x_{\text{tar}}$. At time $t$, an instantaneous discrete flow field is induced by
\begin{displaymath}
    f_t : \*S_t \rightarrow \*S_{t+\epsilon} - \*S_{t},
\end{displaymath}
assuming we have a dense correspondence between the samples on $\surface$. This formulation is similar to prior flow-based deformation techniques \cite{xu2022morig}; however, due to our choice of $\G$, we do not need to explicitly compute point correspondences (see Section \ref{sec:implementation} for details). We deform our mesh by integrating the flow vectors on each mesh vertex. Since the  flow field $f$ is spatially discrete, it must be interpolated to be defined at all vertex locations $\*v \in \*V$. One such method for doing this is by means of a radial basis function (RBF) interpolator \cite{rbfinterpolator}. This method is particularly well-suited for our task because RBF interpolation is grid-free ($f$ does not lie on a regular grid), and it exhibits smooth and stable interpolation for large numbers of points. Flowing each mesh vertex, $\*v^i$, can now be done by integrating the flow field through time
\begin{equation}
    \*v^i_{t+\epsilon} = \*v^i_{t} + \mathrm{RBF}_{f_t}(\*v^i_{t})
\end{equation}
We employ a smoothing RBF interpolator that takes as input a parameter $\lambda$, which controls how well the interpolant fits the displacement vectors $f_t$. An example of our mesh deformations is shown in Figure \ref{fig:exampleDeform}, where we see that interpolations over our energy-minimizing submanifold provide smooth flows on the mesh vertices.

\subsection{Topology-aware deformation subspaces}\label{sec:boundaryRemapping}
Ultimately, we define a smooth space of mesh deformations over a collection of meshes $\M$, akin to the latent space of a generative model. This goal comes with an added challenge: the meshes in $\M$ have varying topology, necessitating a need to discretely switch between topologies while walking through our space. At a high level, we establish a Voronoi partitioning of $\explore$, such that the mesh topology at any point $\*x \in \explore$ is decided by the nearest landmark's corresponding mesh. In the one dimensional case (i.e. with two landmarks and a line connecting them), this is equivalent to switching mesh topologies at the midpoint $t=0.5$.

It is apparent, both qualitatively and quantitatively, that naively switching topologies at $t=0.5$ is suboptimal. We instead compute the optimal switch point, $t^*$, as the time at which the switch incurs minimal change in the deforming mesh. In order to preserve the Voronoi partioning as-is, we transform the boundary conditions on each Delaunay edge to remap $t^*$ to $t=0.5$. Hence, in the final exploration space, the optimal switch points are demarcated simply along Voronoi cells. Quantitative comparisons can be found in Section \ref{sec:results}.

\section{Implementation}\label{sec:implementation}
\begin{figure}
    \centering
    \includegraphics[width=\linewidth]{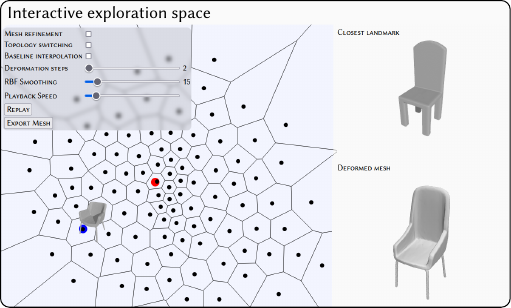}
    \caption{Our interface for real-time shape exploration and deformation, which facilitates interactive exploration of continuous deformations of the landmark meshes. Source and target points are marked by blue and red.}
    \label{fig:GUI}
\end{figure}
All experiments were conducted on a single NVIDIA RTX 3090. We use PyTorch and Adam for all optimization routines \cite{pytorch, adamOptimizer}. Details regarding computation of boundary conditions, discretization of exploration space, training for $\map$, and embedding of landmark meshes can be found in the supplemental material. We additionally implement a graphical user interface that allows one to interact with our explorable deformation space in real time, as seen in Figure \ref{fig:GUI} and included videos.

For our generator $\G$, we employ SP-GAN \cite{li2021spgan}, a state-of-the-art point cloud GAN. SP-GAN is particularly well-suited for our goals because it is capable of representing detailed shapes while implicitly providing a dense correspondence between them, which we utilize in our shape deformation module. SP-GAN takes as input a \emph{latent matrix} which contains a 128-d Gaussian i.i.d vector for each of the 2048 points in the resulting point cloud, thus the generative mapping is $\G : \Z \in \mathbb{R}^{2048 \times 128} \mapsto \X \in \mathbb{R}^{2048\times 3}$.

\begin{figure}
    \centering
    \includegraphics[width=\linewidth, trim=0 10px 0 0]{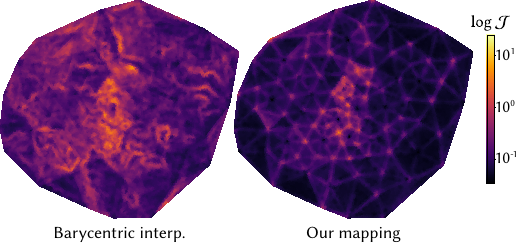}
    \caption{Visualization of the log-energy, as per Equation \ref{eq:finalEnergy}. Naively lifting exploration spaces into the primal domain (e.g. by barycentrically interpolating landmark latents) results in noisy interpolants. Our method incurs far less energy overall, which primarily occurs on the facet boundaries.}
    \label{fig:baryVsOurs}
\end{figure}

\section{Results and Evaluation}\label{sec:results}
We test our method using SP-GAN trained on ShapeNet as our shape generator $\G$ \cite{shapenet}. In particular, we construct three exploration spaces containing chairs, tables, and airplanes---these spaces contain 100, 50, and 25 shapes respectively and are denoted as \emph{chairs-100}, \emph{tables-50}, \emph{airplanes-25}. We demonstrate the flexibility of our exploration space by showing applications such as shape-to-shape deformation (i.e. matching the structure of a target shape with the geometry of a source shape), and free-form exploration of the exploration space (i.e. interpolating in $\explore$ freely without a particular target shape). Our results are compared with ShapeFlow quantitatively and qualitatively; and we similarly use the ShapeNet Manifold dataset \cite{jiang2020shapeflow, shapenetManifold}.

\begin{figure}
    \centering
    \scalebox{-1}[1]{\includegraphics[width=\linewidth,trim=0 128px 0 128px]{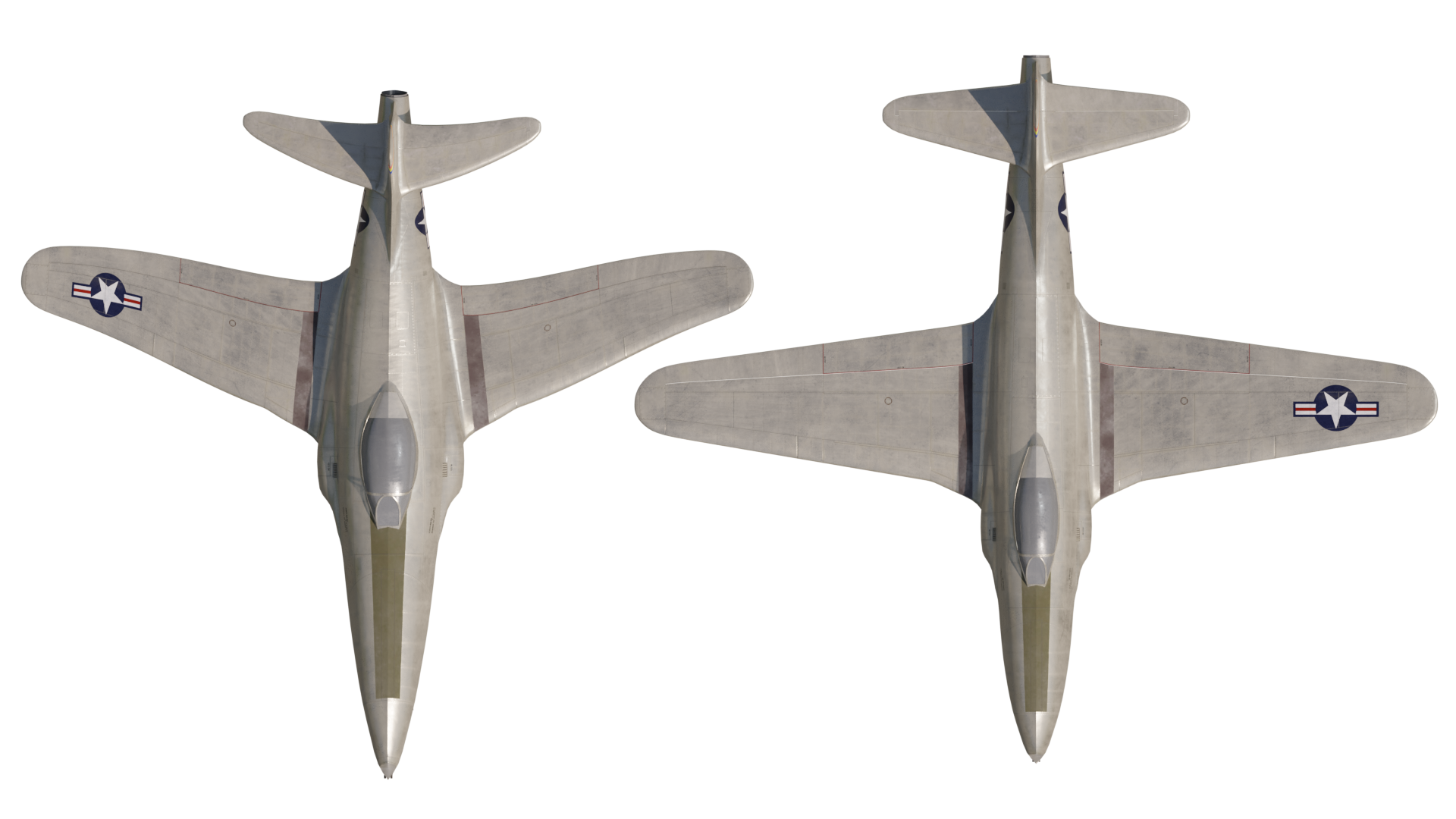}}
    \caption[A]{An out-of-distribution airplane\footnotemark{} (left), unseen by the pre-trained generator, and a resulting mesh from our deformation pipeline (right).}
    \label{fig:outOfDist}
\end{figure}

\paragraph{Shape-to-shape deformation.} 
% \subsubsection{Shape-to-shape deformation} 
Our results on shape-to-shape deformation are presented in Figure \ref{fig:shapeFlowComparison}, whereby source and target landmarks are connected via linear paths in exploration space. We additionally apply our deformation method to an out-of-distribution shape, as seen in Figure \ref{fig:outOfDist}. We compare our results qualitatively and quantitatively with ShapeFlow. The landmark pairs were chosen randomly. We only show comparisons for shapes that are in ShapeFlow's training data (out-of-distribution shapes are not selected). We specifically compare against ShapeFlow's chair model, as other models were unavailable.

\begin{figure*}[p]
    \centering
    \includegraphics[width=0.95\textwidth]{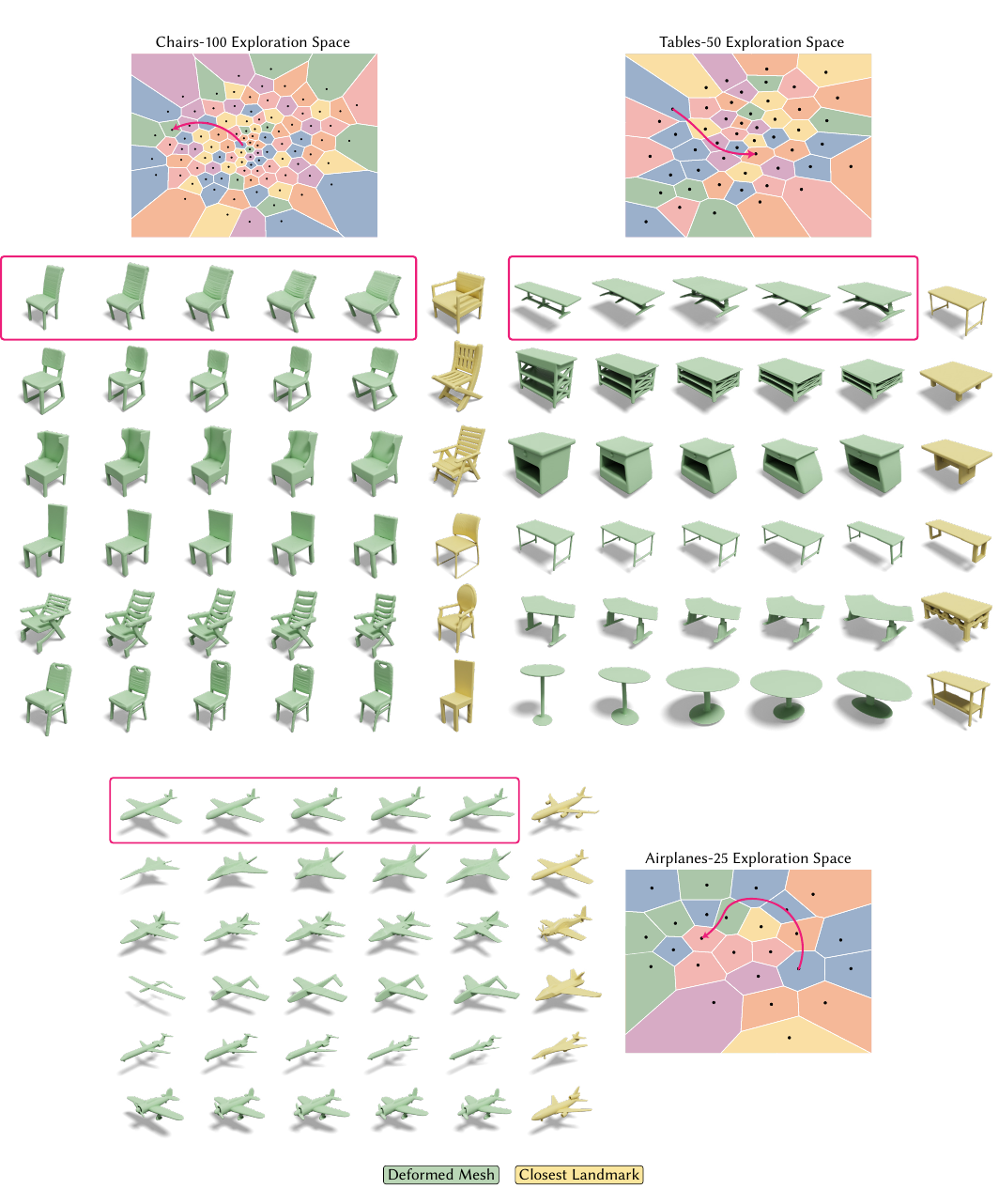}
    \vspace{-1em}
    \caption{Our method also facilitates free-form exploration of the deformation spaces. Here we show example deformations (in green) in our \emph{chairs-100}, \emph{tables-50}, and \emph{airplanes-25} exploration spaces respectively, where times $t=0, 0.25, 0.5, 0.75, 1$ are shown, $t=0$ corresponds to the source mesh. The closest landmark shapes (in yellow) located at the end of the deformation trajectories are shown, along with an example trajectories for the first rows of each category.}
    \label{fig:freeForm}
\end{figure*}

We evaluate our deformations quantitatively by measuring their distributional similarity to the ShapeNet dataset under the Frechet Distance (FD) of a pre-trained PointNet++ classifier's feature space \cite{frechet, qi2017pointnetplusplus}. Table \ref{tab:fidComparison} presents our FD scores alongside ShapeFlow and SP-GAN's meshed outputs (using SAP \cite{Peng2021SAP}). Meshing the point clouds directly leads to geometry that lacks detail and contains artifacts (i.e. holes). We also measure similarity of deformed shapes to their targets under earth mover's distance (EMD). Our deformations are more natural: they preserve the original structure and match the target shapes more closely without significant distortion. ShapeFlow fits the target slightly better under EMD, but this is due to overfitting to a Chamfer distance loss, leading to unnatural surface bowing.
\paragraph{Free-form shape deformation.} We demonstrate the advantage of our two-dimensional exploration space by showing free-form shape deformations, whereby a trajectory starting from a landmark shape can be specified without a particular landmark target. Visualizations of these trajectories can be found in Figure \ref{fig:freeForm}.
\begin{table}[]
\rowcolors{1}{}{lightgray}
    \centering
    \caption{FD and EMD scores of our chair deformations compared with Shapeflow and SP-GAN's meshed outputs. For FD we sample at times $t=0.5,1.0$. We measure EMD between the terminal shape and the target.}
    \begin{tabular}{c | c c}
        Method & FD \{$t=0.5,1.0$\} $\downarrow$ & EMD \{$t=1.0$\} $\downarrow$ \\
        Ours & \textbf{31.6} & 0.082 \\
        ShapeFlow & 78.4 & \textbf{0.069} \\
        SP-GAN (meshed) & 79.2 & - \\
    \end{tabular}
    \label{tab:fidComparison}
\end{table}

\paragraph{Local smoothness of} $\map$. We measure the effect of our mapping on local 1-d trajectories through the exploration space. Specifically, we sample random paths through $\explore$ with lengths proportional to Delaunay facets bounding box dimensions. We compare the energy of our paths to linear paths and optimized geodesics in Table \ref{tab:ablationSmoothness}. Additionally, the Dirichlet energy of our mapping over the entire exploration space is visualized in Figure \ref{fig:baryVsOurs} alongside a baseline.

\begin{table}[]
\rowcolors{1}{}{lightgray}
    \centering
    \caption{Energy of paths from our \emph{chairs-100} exploration space, versus linear interpolations in $\Z$, and optimized geodesics. Our mean energy is smaller than the optimized geodesics, suggesting that the local support provided by our 2-d method gives smoother results than 1-d interpolation in general.}
    \begin{tabular}{r|c c c}
       Metric  & Ours & $\mathcal{Z}$-linear & $\mathcal{Z}$-opt \\
       Mean Energy $\downarrow$& \textbf{1.071} & 1.267 & 1.144 \\
       Max Energy $\downarrow$& 3.741 & 2.796 & \textbf{2.415}
    \end{tabular}
    \label{tab:ablationSmoothness}
\end{table}

\paragraph{Boundary remapping.} We measure the effect of our boundary remapping (Section \ref{sec:boundaryRemapping}). The average Chamfer distance of shapes immediately before and after switching topology are in Table \ref{tab:ablationSwitch}. We see that remapping boundary conditions moderately reduces the geometric distance between shapes at the switch points on average.
\begin{table}[]
\rowcolors{1}{}{lightgray}
    \centering
    \caption{The amount of change incurred by the meshes in \emph{tables-50} exploration space at random switch points is computed. We compare using remapping of optimal switch points, $t^*$, versus no remapping.}
    \begin{tabular}{r|c c}
        Metric & Remapping $t^*$ & w/o Remapping  \\
        Mean CD $\downarrow$& \textbf{262} & 285 \\
        Max CD $\downarrow$& 493 & \textbf{488} \\
    \end{tabular}
    \label{tab:ablationSwitch}
\end{table}

\section{Conclusions and Future Work}\label{sec:conclusion}
In this paper, we presented a new technique for extracting an easily-navigable space for exploring deformations within and between high-quality meshes using a pre-trained 3D shape generative model.
We described an algorithm for computing a smooth 2D parametrization of the generative model's latent space that interpolates a set of given landmark shapes, and we showed how to turn paths along the manifold induced by our parametrization into continuous deformation fields for transferring shape variations from the latent space to high-quality landmark meshes.
We demonstrated our technique by constructing exploration spaces for several shape categories and showed that it produces better interpolations between shapes than an alternative approach, including a recently proposed method for directly learning a deformation space from a large shape collection.

\begin{figure*}[p]
    \centering
    \includegraphics[trim={0 0 0.5cm 0},width=\textwidth]{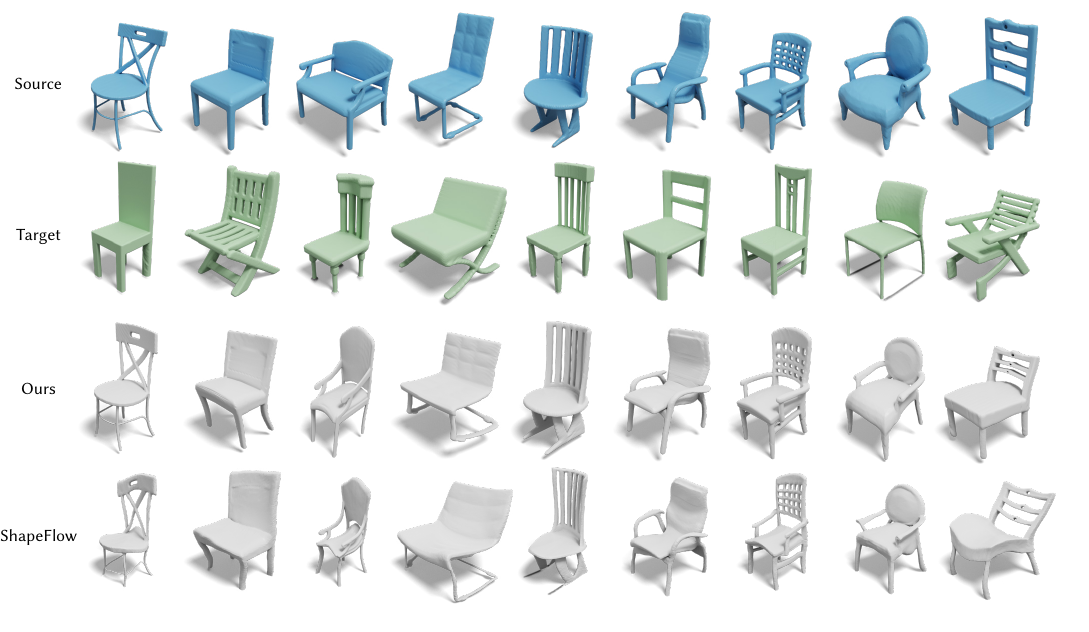}
    \caption{We demonstrate our shape-to-shape deformation results against ShapeFlow, whereby we take random source and target meshes, and deform between them continuously. We visualize the final deformed shapes (at $t=1$) as well as the source and target meshes. We can see that our method exhibits deformations that better preserve the fine details of the original shapes, while matching the structure of the target shapes more closely compared to ShapeFlow.}
    \label{fig:shapeFlowComparison}
\end{figure*}

\paragraph{Limitations.} Our method is not without limitations.
For instance, switching between meshes can result in visual discontinuities, especially if the landmarks are tightly packed.
One way to alleviate this issue would be to produce more gradual topology changes between the two meshes, rather than a single discrete switch: for example, by gradually replacing parts of the source mesh~\cite{Jain:2012:3DModelRecombination,NeuralTemplate}. Additionally, like other flow-based deformation methods, our method may produce warped geometry, particularly when `rigid' parts are deformed non-rigidly (e.g. column 3 in Figure \ref{fig:shapeFlowComparison}). Imposing regularization energies (e.g. an ARAP energy \cite{ARAP, ARAPReg}) on the induced flow field is a potential avenue for improvement. 
Finally, while our method supports real-time interaction, the training remains time consuming. Amortizing the training of $\map$ by \emph{learning to predict} an exploration space given a set of landmarks is of much interest.
Such a system could facilitate powerful forms of interaction: e.g., `coarse-to-fine' exploration, in which a user starts with an exploration space over a large set of shapes, then iteratively drills down into spaces of smaller subsets of shapes. Currently our method cannot easily accommodate this interaction paradigm.

\paragraph{Future work.} One immediate direction for future work is to apply our method to other latent variable generative models.
The challenge here is in producing deformation fields when there does not exist a dense correspondence in $\G$'s output space (as in SP-GAN).
Also, given their recent popularity, it would be interesting to apply our method to diffusion models.
The challenge here is the expensive computation of Jacobians that are necessary for training the map $\Phi$.
One possibility is to train a surrogate forward-pass model that approximates multiple diffusion steps \cite{diffusionGAN}.

It would also be interesting to apply the subcomponents of our method to different applications or domains.
For instance, our procedure for building $\explore$ is not specific to 3D shape latent spaces; one could imagine using it to construct easily-navigable exploration spaces for other visual data domains, such as image collections.
\footnotetext{Airplane model from CGTrader user Boba3D.}

%%
%% The acknowledgments section is defined using the "acks" environment
%% (and NOT an unnumbered section). This ensures the proper
%% identification of the section in the article metadata, and the
%% consistent spelling of the heading.
\begin{acks}
This material is based upon work supported by the National Science Foundation Graduate Research Fellowship under Grant No. 2040433.
\end{acks}

%%
%% The next two lines define the bibliography style to be used, and
%% the bibliography file.
\bibliographystyle{ACM-Reference-Format}
\bibliography{mybib}

\begin{figure*}[p]
    \centering
    \includegraphics[trim={0 0.5cm 0.5cm 0},width=\textwidth]{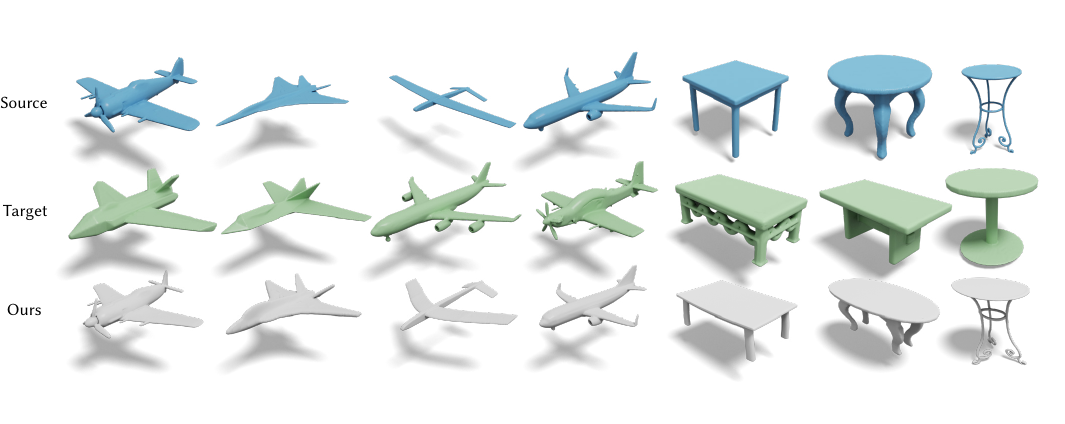}
    \caption{Additional shape-to-shape deformation results from our \textit{airplanes-25} and \textit{tables-50} exploration spaces. Our method is able to capture complex deformations; for instance, in column four we see the airliner's wings bending forward to match the straight wing's of the propeller plane. Additionally, in the last column we see that our method is able to locally deform the table top while maintaining the geometry elsewhere.}
    \label{fig:shapeToShape}
\end{figure*}

%%
%% If your work has an appendix, this is the place to put it.
\appendix

%\section{stuff}

\end{document}

\title{Supplemental Material: Explorable Mesh Deformation Subspaces from Unstructured 3D Generative Models}
%%
%% By default, the full list of authors will be used in the page
%% headers. Often, this list is too long, and will overlap
%% other information printed in the page headers. This command allows
%% the author to define a more concise list
%% of authors' names for this purpose.
%%
%% The abstract is a short summary of the work to be presented in the
%% article.
% \begin{abstract}
% \input{sec_abstract.tex}
% \end{abstract}

% \received{20 February 2007}
% \received[revised]{12 March 2009}
% \received[accepted]{5 June 2009}

%%
%% This command processes the author and affiliation and title
%% information and builds the first part of the formatted document.
\maketitle

\appendix

\section{Experimental and Implementation Details}

\subsection{Latent code optimization}
\begin{figure*}
    \centering
    \includegraphics[width=\textwidth, trim = 0 10mm 0 0]{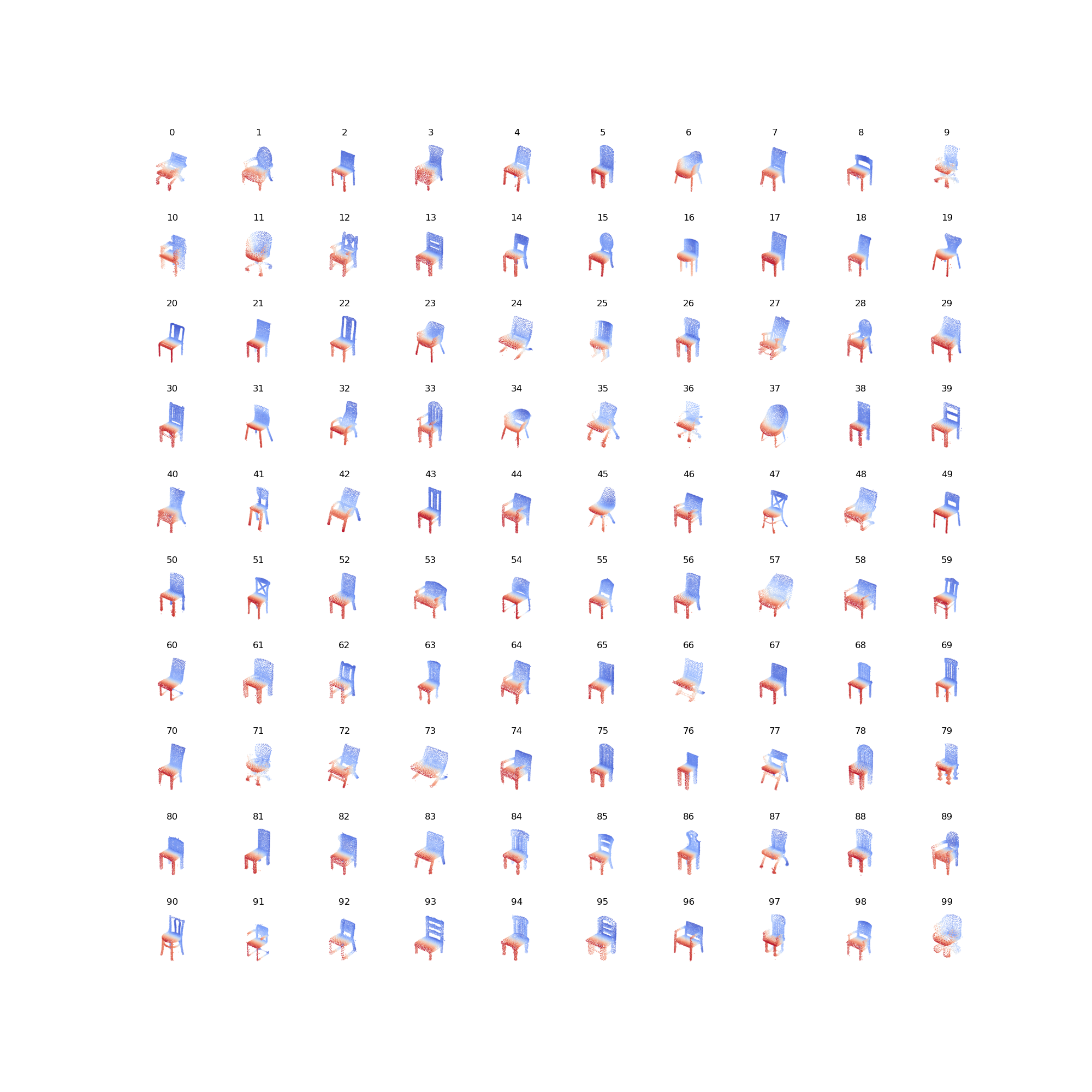}
    \caption{Point cloud visualizations of the chairs in \emph{chairs-100} after being projected into SP-GAN's latent space.}
    \label{fig:chairsPc}
\end{figure*}

\begin{figure*}
    \centering
    \includegraphics[width=\textwidth, trim = 2cm 10mm 2cm 0]{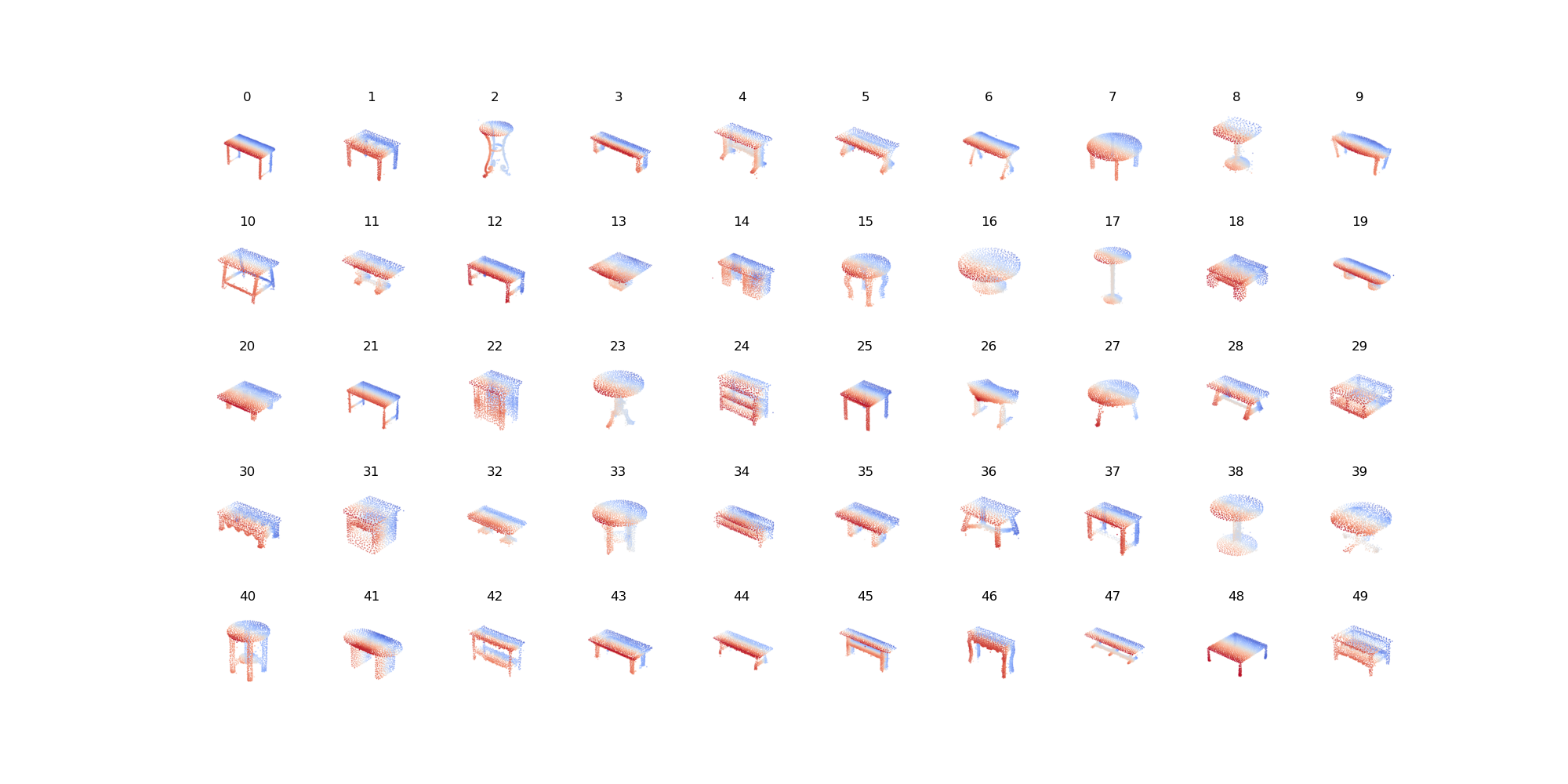}
    \caption{Point cloud visualizations of the tables in \emph{tables-t0} after being projected into SP-GAN's latent space.}
    \label{fig:tablesPc}
\end{figure*}

\begin{figure*}
    \centering
    \includegraphics[width=0.6\textwidth, trim = 0 10mm 0 0]{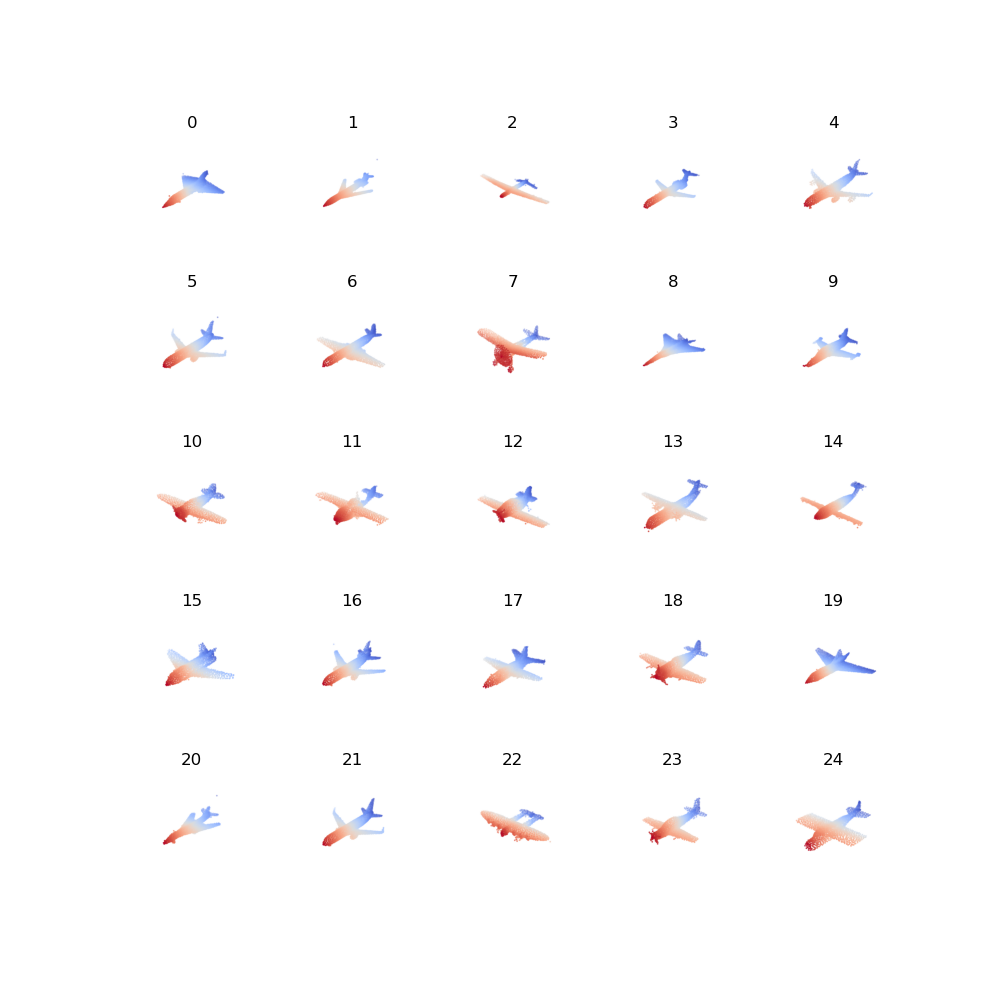}
    \caption{Point cloud visualizations of the airplanes in \emph{airplanes-25} after being projected into SP-GAN's latent space.}
    \label{fig:airplanesPc}
\end{figure*}

The given meshes in $\meshes$ are projected into $\G$'s latent space by minimizing the Chamfer distance of a random point sampling of each mesh, $\*P$, to the output of the generator with respect to a latent code $\*z_m$, making the objective $\argmin_{\*z_m} \mathrm{CD}(\G(\*z_m), \*P)$. This optimization is done in a coarse-to-fine manner, where the sampling progressively gets denser; starting from $2^{11}$ and growing to $2^{15}$ points, doubling every 800 iterations. We used a point sampling schedule of $2048, 4069, 8192, 16384, 32768$, resampling every 800 iterations. The coarse-to-fine procedure was critical for avoiding local minima in the optimization. We visualize all projections of our shapes in each exploration space in Figures \ref{fig:chairsPc}, \ref{fig:tablesPc}, and \ref{fig:airplanesPc}.

\subsection{Embedding of meshes into $\explore$}
The K nearest neighbor graph used to embed our landmarks into $\explore$ uses $k=5$ and the similarity between landmarks is determined by the dense correspondence distance given by SP-GAN's resulting point clouds. That is, the distance metric is the sum of Euclidean distances between points in $\G(\*z_1)$ and $\G(\*z_2)$ for two landmarks $\*z_1$, $\*z_2$. 

We embed the input meshes $\meshes$ into the exploration space via a two-stage optimization process. First we optimize the embedding $\*X = \{\*x_1, ..., \*x_N\}$ using the triplet loss formulation given in Section 4.4, after 600 optimization iterations, we compute a Delaunay triangulation, $\mathcal{T}$, of $\*X$. With this triangulation, we begin to impose $\ell_2$ losses on the areas of triangles in $\mathcal{T}$ as well as the minimum interior angles of each triangle. In a sense, this process iteratively makes the embedding ``more Delaunay.'' This process alone may cause the embedding to shrink to a single point, hence, we impose a constraint by ``pinning'' the convex hull of $\*X$ before beginning stage two---these points do not get updated during optimization. In order to prevent the triangulation from becoming entangled (i.e. overlapping edges), we use a much smaller learning rate in the second stage of this process. We use Adam optimizer with a learning rate of $0.1$ for the first stage, and $0.005$ for the second stage. Finally, we execute an adjustment step that snaps points in $\*X$ to the nearest point on the convex hull, if they are within a distance threshold from the hull. This prevents large sliver triangles from forming along the hull.

Note that alternative methods for the first stage can act as a drop-in replacement. That is, one could use, say, t-SNE or UMAP for the first stage of this pipeline (the second stage is unaffected).

\subsection{Discretization of $\mathcal{E}$}
We discretize the exploration space using the CAL-FEM Python package. Each facet in the Delaunay triangulation is densely subdivided into uniform elements, and the number of elements is determined by the area of the facet. The exploration spaces \emph{chairs-100}, \emph{tables-50}, and \emph{airplanes-25} were discretized into approximately $18\cdot10^3, 13\cdot10^3$ and $15\cdot10^3$ facets each.

\subsection{Computation of boundary conditions}
Similar to the procedure suggested by [Laine 2018], we discretize the geodesics paths into polylines, and the objective is then given by 
\begin{displaymath}
\textstyle
\argmin_{\*z_1, ..., \*z_{n-1}}\sum_{i=0}^n \|\G(\*z_i) - \G(\*z_{i+1})\|^2    
\end{displaymath}
where $\*z_{0...n}$ are nodes along a polyline with $n$ nodes. We employ a coarse-to-fine optimization of our polylines in $\Z$. The polylines are initialized as straight lines with 8 nodes connecting the source and target latent codes. The polyline is subdivided every 100 iterations, doubling the number of nodes until the polyline has 64 nodes.

\subsection{Computing switch points \& remapping boundary conditions}
Our boundary conditions are given as polylines with 64 nodes that connect latents $\*z_i$ to $\*z_j$. In order to compute the point in which we switch from mesh $\*M_i$ to mesh $\*M_j$, we first compute two deformation sequences: one from $\*M_i$ to $\*M_j$, and the other $\*M_j$ to $\*M_i$. Given these two sequences of meshes, we can identify the time $t^*$ where the chamfer distance between the meshes is minimal. We find the optimal switch point in a subsection of the deformation sequence centered around $t=0.5$, i.e. we do not take a switch point to be, say, $t=0.01$, rather we only consider $t$ values in $[0.35, 0.65]$. This is to prevent overly distorting the boundary conditions. 

Remapping the polyline is done by dilating both sides of the polyline such that $t^*$ is mapped exactly to $t=0.5$. Hence all of the switch points' boundaries can be visualized by a standard Voronoi diagram.

\subsection{The energy-minimizing map $\map$}

\textit{Training.} Our map $\map$ from exploration space to $\Z$ is implemented as a four layer multi-layer perceptron with sinusoidal positional encoding [Mildenhall et al. 2020], where $L=5$
\begin{displaymath}
\gamma(\*x) = \left(\sin(2^0 \pi \*x), \cos(2^0 \pi \*x), ..., \sin(2^L \pi \*x), \cos(2^L \pi \*x)\right).
\end{displaymath}
The MLP is optimized via stochastic gradient descent with a learning rate of $3\cdot10^{-4}$ as per the objective in Equation 2. We use batch sizes of 256 with gradient accumulation. Our training times vary from 12 to 24 hours for our smallest and largest exploration spaces.

\textit{Inference.} Our formulation of the submanifold objective (Section 4.2) assumes that $\Z$ is piecewise linear, hence in order to do inference on $\map$ properly, we must query it in a similar fashion. At inference time, a point in exploration space $\*x \in \explore$ is decoded into primal space by first identifying which FEM facet $\*x$ resides in. Then we compute a barycentric interpolation of the functional values of the facet's vertices lifted into primal space. Additionally, if any of these vertices lie on our boundary, we ``swap'' out the functional value with the boundary value, which is the same procedure that we use during training. 

\subsection{Deformation module}
For a given path connecting source and target points in exploration space, we accumulate the vertex displacements via the discrete Euler method in Equation 4. For all rendered results in the paper, we integrate over 180 samples along the given path. We employ a custom GPU-accelerated smoothing radial basis function interpolator that has a throughput of 72 steps per second, which facilitates real-time interaction. The smoothing parameter is chosen for each exploration space independently: 15, 50, 15 for the \emph{airplanes-25}, \emph{tables-50}, \emph{chairs-100} spaces respectively. Preliminaries for RBF interpolation are in Appendix B.

\subsection{Obtaining shapes for exploration spaces}
Our exploration spaces: \emph{chairs-100}, \emph{tables-50}, \emph{airplanes-25} were populated using a semi-random shape retrieval routine. We start with a set of $N$ hand-picked shapes ($N=5$ in our case), then iteratively select shapes from the dataset based on similarity. In particular, at each step in this process we sample a random shape from the top-$K$ nearest shapes in ShapeNet. We vary the $K$ parameter across experiments to accommodate for lack of diversity in certain shape categories. In particular, we used $K=2000$, $K=1500$, and $K=5000$ for chairs, tables, and airplanes respectively. A temperature parameter $\gamma$ was used to control a similarity-weighted softmax over the retrieved $K$ shapes. Finally, the distance metric used to measure shape similarity is the dense-correspondence distance given by SP-GAN, as mentioned in Section 6.

\section{RBF Interpolation Preliminaries}
Given a set of data samples $f(\*x_0), \dots, f(\*x_n)$ and \emph{radial basis functions} centered those samples, $\phi(\*x_k) = \hat{\phi}(\|\*x - \*x_k\|)$, RBF interpolation represents an interpolant as a weighted combination of the bases
\begin{align*}
S(\*x) &= \sum_{i=0}^n w_i \cdot \phi(\|\*x - \*x_i\|),\\
\mathrm{s.t.}&\quad S(\*x_i) = f(\*x_i)\, \forall i
\end{align*}
whose weights can be solved via the linear system $\*A \*w = \*b$, where
\begin{displaymath}
    \begin{bmatrix}
    \myphi{0}{0} & \dots & \myphi{n}{0} \\
    \myphi{0}{1} & \dots & \myphi{n}{1} \\
    \vdots & \ddots & \vdots \\
    \myphi{0}{n} & \dots & \myphi{n}{n}
    \end{bmatrix}\begin{bmatrix}
        w_0 \\ w_1 \\ \vdots \\ w_n
    \end{bmatrix}=\begin{bmatrix}
        f(\*x_0) \\ f(\*x_1) \\ \vdots \\ f(\*x_n)
    \end{bmatrix}
\end{displaymath}
A common choice for the radial function is the \emph{thin plate spline}, which is given by $\phi(r)=r^2 \ln r$, and is our choice of basis function.

The smoothing variant of RBF interpolation requires that we add a polynomial term to the interpolant. That is, we add to $S(\*x)$ a sum of weighted monomials up to a specified degree, and require that the interpolant evaluates to a be \emph{exactly} polynomial if the data itself comes from a polynomial (this additional condition is necessary to specify a unique solution). The linear system eventually becomes
\begin{displaymath}
    \begin{bmatrix}
        \*A + \lambda \*I & \*P \\
        \*P^T & \*0
    \end{bmatrix}
    \begin{bmatrix}
        \*w \\ \*d
    \end{bmatrix}
    =
    \begin{bmatrix}
        \*b \\ \*0
    \end{bmatrix}
\end{displaymath}
where $\*P$ is a matrix of monomial terms and $\*d$ is their coefficients. Finally, the smoothing parameter $\lambda$ (which is mentioned in our method), is incorporated by adding a constant term to the diagonal of $\*A$. For further reading, please refer to [Anjyo et al. 2014].

